\documentclass[twocolumn,aps,nofootinbib,amsfonts,superscriptaddress]{revtex4}
\usepackage{amsmath}
\usepackage{epsfig,epsf}
\usepackage{graphicx}
\usepackage{verbatim}
\usepackage{epstopdf}

\def\beq{\begin{equation}}
\def\eeq{\end{equation}}
\def\bea{\begin{eqnarray}}
\def\eea{\end{eqnarray}}
\def\eq#1{{Eq.~(\ref{#1})}}
\def\fig#1{{Fig.~\ref{#1}}}

\newcommand{\as}{\alpha_s}
\newcommand{\Lb}{\left(}
\newcommand{\Rb}{\right)}

\begin{document}

\voffset1.5cm
\title{Hadron multiplicity in pp and AA collisions at LHC from the Color Glass Condensate  }
\author{Eugene Levin}
\affiliation{
 Departamento de F\'\i sica, Universidad T\'ecnica
Federico Santa Mar\'\i a, Avda. Espa\~na 1680,
Casilla 110-V,  Valparaiso, Chile}
\affiliation{Department of Particle Physics,  Tel Aviv University, Tel Aviv 69978, Israel}
 \author{Amir H. Rezaeian}
\affiliation{
 Departamento de F\'\i sica, Universidad T\'ecnica
Federico Santa Mar\'\i a, Avda. Espa\~na 1680,
Casilla 110-V,  Valparaiso, Chile}
\date{\today}
\begin{abstract}
We provide quantitative predictions for the rapidity, centrality and energy
dependencies of inclusive charged-hadron productions for the forthcoming LHC measurements in
nucleus-nucleus collisions based on the idea of gluon saturation in the
color-glass condensate framework. Our formulation gives very good
descriptions of the first data from the LHC for the inclusive charged-hadron
production in proton-proton collisions, the deep inelastic scattering at HERA at small Bjorken-$x$, and the hadron
multiplicities in nucleus-nucleus collisions at RHIC.

\end{abstract}
\maketitle


\section{Introduction}

It has been widely discussed that QCD predicts at high-energy formation of a new
state of matter, the so-called color-glass condensate (CGC) 
\cite{HDT}. In this state, the density of quarks and gluons $\rho$ with transverse momenta less that $Q_s$ reaches a high value,
 namely $\rho \propto 1/\as(Q_s) \,\,\gg\,\,1$  where $\as$ is the strong coupling
constant and $Q_s$ is a new momentum scale (saturation momentum) that
increases with energy. Therefore, $\as\Lb Q_s\Rb \,\,\ll\,\,1$ and
this fact allows us to treat this system on solid theoretical
basis. One of the most characteristic and qualitative
consequences of the CGC is the emergence of a new mechanism for hadron production at
high energy. In this approach, the process of secondary hadron
production goes in two stages: production of gluon mini-jet with a
typical transverse momentum $Q_s$; and the decay of gluon mini-jets
into hadrons.

The CGC picture for the hadron production has passed two critical
tests. First, it explains the main features of hadron
multiplicity in heavy ion-ion collisions at RHIC (KLN papers \cite{KLN}); and it
predicts the inclusive hadron production in proton-proton ($pp$) collisions
\cite{LR} at the LHC which was recently confirmed experimentally by the CMS
collaboration at $\sqrt{s}=7$ TeV \cite{CMS} and the lower LHC
energies \cite{CMS2,ALICE,ATLAS}, see also Ref.~\cite{MCLPR}. The
ion-ion collisions at the LHC will lead to a crucial test of the CGC
approach. Indeed, the KLN success \cite{KLN} in the description of
RHIC data shows that we are dealing with scatterings of dense
partonic systems even at rather low energies. At higher energies we
expect that the density of scattering systems increases resulting in
more transparent and unambiguous evidence for the creation of the CGC.

In this paper we wish to extend our approach \cite{LR} to ion-ion
($AA$) collisions at the LHC.  As it was discussed in Ref.~\cite{LR},
we improve the KLN approach \cite{KLN} in various ways incorporating
the unintegrated gluon density that describes HERA data at small
Bjorken-$x$. We also include our knowledge on proton-proton
scatterings at the LHC to make our prediction more reliable. In
particular, we re-calculate the saturation momentum for nuclei from
the saturation momentum in the proton target obtained from DIS data at
HERA and confront it with the RHIC's gold-gold data. On the contrary, in
the KLN approach \cite{KLNLHC}, the LHC saturation momentum was found
via an extrapolation of the energy dependence of the saturation scale
at RHIC in the BFKL region.

In the next section, we discuss the $k_T$-factorization and our main
formulation for the inclusive hadron production in $pp$ and $AA$
collisions. In particular, we introduce the impact-parameter dependent
saturation scale for a proton and a nuclear target. Section III is
devoted to comparison with the experimental data and to discussion of
various predictions for the LHC energies.  Finally, we conclude in
Sec. IV.

\section{Main formulation}

The gluon jet production in nucleus-nucleus  collisions can be described by $k_T$-factorization given by \cite{KTINC},
\beq \label{M1}
\frac{d \sigma}{d y \,d^2 p_{T}}=\frac{2\alpha_s}{C_F}\frac{1}{p^2_T}\int d^2 \vec k_{T} \phi^{G}_{A}\Lb x_1;\vec{k}_T\Rb \phi^{G}_{A}\Lb x_2;\vec{p}_T -\vec{k}_T\Rb,
\eeq
where $x_{1,2}=(p_T/\sqrt{s})e^{\pm y}$, $p_T$ and $y$ are the
transverse-momentum and rapidity of the produced gluon
mini-jet. $\phi^{G}_{A}(x_i;\vec k_T)$ denotes the unintegrated gluon
density and is the probability to find a gluon that
carries $x_i$ fraction of energy with $k_T$ transverse momentum in the
projectile (or target) nucleus $A$. We defined $C_F=(N_c^2-1)/2N_c$
where $N_c$ denotes the number of colors.

The $k_T$-factorization has been proven \cite{KTINC} for the
scattering of a diluted system of parton with a dense one. Such a
process can be characterized by two hard scales: the transverse
momentum of produced particle $p_T$ and a saturation scale which are
both larger than the soft interaction scale $\mu$. The ion-ion
scattering is a typical example in which we have three scales: $p_T$
and two saturation scales for the projectile and the target. The $k_T$-factorization
might be then violated only at low $x_1$ and $x_2$, and when $p_t$ is
smaller than both saturation scales. Outside of this kinematic region
we are actually dealing with scatterings of diluted-dense system since
one of the saturation scales is small.  For the case of scatterings of dense-dense
system of partons, the $k_T$-factorization has not yet been
proven. Nevertheless, the success of the KLN approach \cite{KLN} which is based on
the $k_T$-factorization, in description of the experimental data at RHIC for gold-gold collisions suggests that the $k_T$-factorization is currently the
best tool that we have at our disposal for the processes of ion-ion
scatterings.

The relation between  the unintegrated gluon density $\phi^G_{A}$ and the
color dipole-nucleus forward scattering amplitude has been obtained in
Ref.~\cite{KTINC}. It reads as follows
\beq \label{M2}
\phi^G_A\Lb x_i;\vec{k}_T\Rb=\frac{1}{\alpha_s} \frac{C_F}{(2 \pi)^3}\int d^2 \vec b d^2 \vec r_T
e^{i \vec{k}_T\cdot \vec{r}_T}\nabla^2_T N^G_A\Lb x_i; r_T; b\Rb,
\eeq
with notation
\beq \label{M3}
N^G_A\Lb x_i; r_T; b \Rb =2 N_A\Lb x_i; r_T; b \Rb - N^2_A\Lb x_i; r_T; b \Rb,
\eeq
where $N_A\Lb x_i; r_T; b \Rb$ is the dipole-nucleus forward scattering amplitude which satisfies 
the perturbative nonlinear small-$x$ Balitsky-Kovchegov (BK) quantum evolution equation \cite{bk}.
In the above, $r_T$ denotes the dipole transverse size and $\vec b$ is the impact parameter of the scattering.

Substituting  \eq{M2} into \eq{M1}, and after analytically performing some integrals, we obtain \cite{KTINC,LR},
\bea \label{M4}
&&\frac{d \sigma(y;p_T;\overline{B})}{d y \,d^2 p_{T}} =
\frac{2C_F\alpha_s(p_T)}{ (2\pi)^4} \int_{B_1}^{B2} d^2 \vec B \int
d^2 \vec b d^2 \vec r_T e^{i \vec{p}_T\cdot \vec{r}_T} \nonumber\\ &&
\hspace{1.5cm}\times \frac{ \nabla^2_T N^G_A\Lb x_1; r_T; b\Rb
\nabla^2_T N^G_A\Lb x_2; r_T; b_{-} \Rb
}{p^2_T\alpha_s\left(Q_A\left(x_1;b\right)\right)\alpha_s\left(Q_A\left(x_2;b_{-}\right)\right)
}, \
\eea 
where $\vec B$ is the impact parameter between the center of two
nuclei, $\vec b$ and $\vec{b}_{-}=\vec b-\vec{B}$ are the impact parameter
between the interacting nucleons with respect to the center of two
nuclei. We will see below that we can neglect the impact parameter of
the produced gluon-jet from the center of the nucleons.

 In the above, we extended the $k_T$-factorization given in
\eq{M1} by introducing a running strong-coupling $\alpha_s$. For the running $\alpha_s$ we employ the same prescription used in Ref.~\cite{LR} for $pp$
collisions.  The saturation scale $Q_A(x_{1,2},b)$ depends on
the $x_{1,2}$ and the impact parameter and will be introduced in the
following. A given centrality bin corresponds to ranges of the
impact-parameter $\overline{B}\in[B_1,B_2]$ of the collisions. The
procedure how to select the event with a fixed $\overline{B}$ and its
relation to the centrality bins is well-known, see for example
Refs.~\cite{KLNLHC,WIED}.

Notice that the relation between the unintegrated gluon density and the
forward dipole-nucleus amplitude Eqs.~(\ref{M2},\ref{M3}) in the
$k_T$-factorization Eq.~(\ref{M1}) is not a simple Fourier
transformation which is commonly used in literature and also depends on
the impact-parameter. The impact-parameter dependence in these
equations is not trivial and should not be in principle assumed as an
over-all factor. Using the general properties of high density QCD that
the underlying physics depends only on the saturation scale we
reconstruct the dipole-nucleus scattering amplitude in two
steps. First, we choose the saturation model which effectively
incorporates all known saturation properties \cite{LR} driven by the
BK equation including the impact-parameter dependence of the dipole
amplitude \cite{LTSOL}. This model describes both the HERA DIS data at
small-$x$  \cite{WAKO} and the proton-proton LHC data \cite{LR}.  Second, we
replace the proton saturation momentum by that of the nucleus. The dipole-nucleus scattering amplitude in our model is given by
\bea \label{M5}
N_A\Lb x; r; b\Rb=\left\{\begin{array}{l} N_0\,\Lb
\frac{\mathcal{Z} }{2}\Rb^{2 (\gamma_s\,\,+\,\,\frac{1}{\kappa \lambda
Y}\ln\Lb\frac{2}{\mathcal{Z}
}\Rb)}\,\,\,\,\,\mbox{for}\,\,\mathcal{Z}\leq 2;\\
\\ 1-\exp\Lb -\mathcal{A} \ln^2\Lb \mathcal{B}
\mathcal{Z}\Rb\Rb\,\,\,\,\,\,\,\,\,\,\mbox{for}\,\,\mathcal{Z} >2;\end{array}
\right.
\eea
where we defined $\mathcal{Z}=r\, Q_A(x;b)$, $Y=\ln(1/x)$ and $\kappa
= \chi''(\gamma_s)/\chi'(\gamma_s)$ where $\chi$ is the LO BFKL
characteristic function.  The parameters $\mathcal{A}$ and
$\mathcal{B}$ are determined uniquely from the matching of $N_A$ and
its logarithmic derivatives at $\mathcal{Z}=2$. The nucleus
saturation scale is given by
\beq \label{M8}
Q^2_{A}\Lb x;b \Rb\,\,\,=\,\,\, \int d^2 \vec b'~T_A\Lb\vec{b} - \vec{b}' \Rb\,Q^2_p\Lb x; b'\Rb.
\eeq
In the above, the proton saturation scale $Q_p$ is defined as
\beq \label{M6}
Q_{p}(x;b')\,\,=\,\,\Lb \frac{x_0}{x}\Rb^{\frac{\lambda}{2}}\,\exp\left\{- \frac{b'^2}{4 (1 - \gamma_{cr}) B_{CGC}}\right\}.
\eeq
and $T_A\Lb B\Rb$ denotes the nuclear thickness. We use for the
nuclear thickness the Wood-Saxon parametrization \cite{WS}.  Notice
that for small $\mathcal{Z}\le 2$, the effective anomalous dimension
$\gamma_{cr}=1-\gamma_s$ in the exponent in the upper line of
Eq.~(\ref{M5}) rises from the BK value towards the DGLAP value.  
The ansatz given in Eq.~(\ref{M5}) for a proton target was first
introduced in Ref.~\cite{o-cgc}. The parameters $ \lambda, \gamma_s,
N_0$, $x_0$ and $B_{CGC}$ are obtained from a fit to the DIS data at
low Bjorken-$x$ $x<0.01$ with a very good $\chi^2/\text{d.o.f.}=0.92$
\cite{WAKO}.

\begin{figure}[!t]
             \includegraphics[width=8 cm] {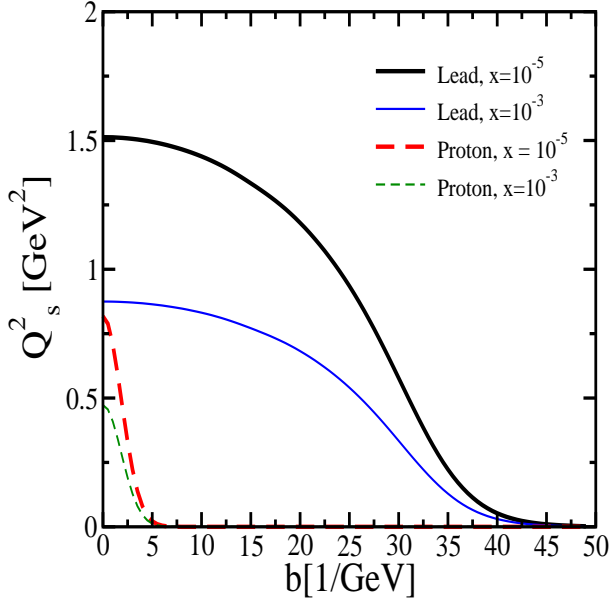}
             \caption{The impact parameter dependence of the saturation scale for proton and lead at $x = 10^{-3}$ and $x=10^{-5}$.\label{f1}}       
\end{figure}

The saturation
scale $Q_A$ in nuclei is proportional to the density of partons in the
transverse plane (see for example Refs.~\cite{HDT,KLN,QSA}) and
\eq{M8} takes this fact into account.  In \eq{M8}, $\vec{b}'$ is the
impact parameter of the color dipole with respect to the center of the
proton, $\vec{b} $ is the impact parameter of the nucleon with respect
to the center of the nucleus and $\vec{b}-\vec{b}'$ is, therefore, the position of the dipole with respect to the center of
the nucleon. The nuclear saturation scale defined in Eq.~(\ref{M8}) gives
$Q_A^2\approx Q_p^2 A^{1/3}$, in agreement with the basic idea of
saturation and the CGC picture
\cite{HDT,KLN,QSA}. This is because $T_A\sim A^{1/3}$ (where A is the
effective mass number of the nucleus in a given centrality) and the
integral in Eq.~(\ref{M8}) is approximately over the nucleon
size. Note that since here the nucleon saturation scale depends on
impact-parameter, it is not then correct to define the saturation scale on
a nuclear target with a simple prefactor $A^{1/3}$ scaling. In \fig{f1}
we show that the impact-parameter dependence of the saturation scale
for proton and lead obtained from Eqs.~(\ref{M8},\ref{M6}) are quite
different as we expected.  \fig{f1} demonstrates the well-known
features of nucleus scattering, namely the typical $b'$ in \eq{M8} is
much smaller than the typical $b \sim R_A$ ($R_A$ is the nuclear radius) in the nuclear thickness.

In order to simulate the
behavior of gluon density at large $x\to 1$, we product the
unintegrated gluon density with $(1-x)^4$ as prescribed by quark
counting rules and the HERA data on DIS at large-$x$ \cite{LR}. Notice that the
contribution of $(1-x)^4$ correction at $\sqrt{s}=5.5$ TeV for $AA$ collisions and $\eta \leq 4.5$ is
less than $4\%$, and for $4.5<\eta<6$ is less than $13\%$. Therefore,
the main contribution of the unintegrated gluon density at the LHC
high-energy in the kinematic region considered here for both $AA$ and also $pp$ collisions \cite{LR} comes from the
small-$x$ region where the saturation physics is important. For the importance of the saturation effects in $pA$ collisions at the LHC, see Ref.~\cite{RS}.

The $k_T$-factorization Eqs.~(\ref{M1},\ref{M4}) gives the
cross-section of radiated gluon mini-jets with zero mass while what is
actually measured experimentally is the distribution of final hadrons.
We therefore should model the nonperturbative hadronization stage of gluon
mini-jets.  We employ the Local
Parton-Hadron Duality principle \cite{LPHD}, namely the
hadronization is a soft process and cannot change the direction of
the emitted radiation. Hence
the form of the rapidity distribution of the mini-jet and the produced hadron are different only with a  
numerical factor $\mathcal{C}$.  It is well-known that the general
assumption about hadronization leads to the appearance of mass of the
mini-jet which is approximately on average equal to $m^2_{jet} \simeq 2 \mu \langle p_T\rangle$ \cite{KLN,LR} where $\mu$ is the scale of soft
interaction. The mini-jet mass $m_{jet}$ effectively incorporates the
nonperturbative soft pre-hadronization in the pseudo-rapidity space and can be approximately related to the saturation scale \cite{LR}.
Accordingly, one should also correct the kinematics every where in
Eq.~(\ref{M4}) due to the presence of a non-zero mini-jet mass,
namely replacing $p_T\to
\sqrt{p^2_T+ m^2_{jet}}$ in $x_1, x_2$ and also in the denominator of
$1/p_T^2$. Finally, in order to take account of the difference between
rapidity $y$ and the measured pseudo-rapidity $\eta$, we employ the Jacobian transformation between $y$ and $\eta$ \cite{LR}.

Following Ref.~\cite{LR}, in the spirit of the geometrical-scaling property of the scattering amplitude,
we obtain the charged-particle multiplicity distribution at a
fixed centrality but various energies from the corresponding mini-jet
cross-section Eq.~(\ref{M4}) divided by the average area of interaction $\sigma_{s}=\mathcal{M} \pi \Big \langle \vec b^2_{jet} \Big\rangle=
\mathcal{M} \pi \Big \langle b^2+b^2_{-} \Big\rangle$. In a similar way, in order to obtain   
 the charged-particle multiplicity distribution at various centrality
 bins but a fixed energy, we divide the mini-jet cross-section
 integrated in ranges of $\overline{B}\in[B_1, B_2]$ with the
 corresponding relative interaction area $\sigma_{s}=\mathcal{M} \pi
 (B_2^2-B_1^2)$.  

\section{Discussion and predictions}

\begin{figure}[!t]
             \includegraphics[width=8 cm] {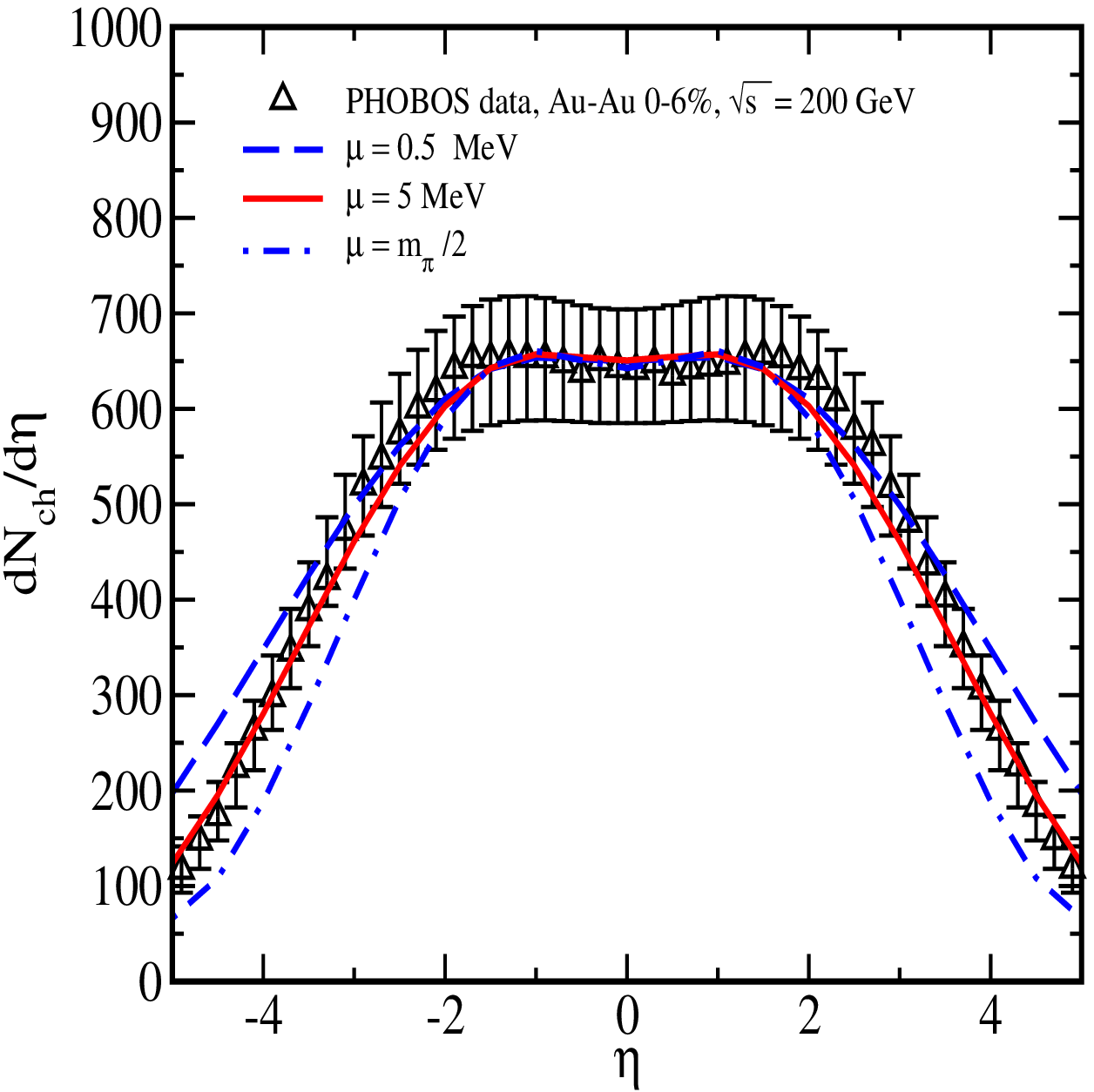}
              \includegraphics[width=8 cm] {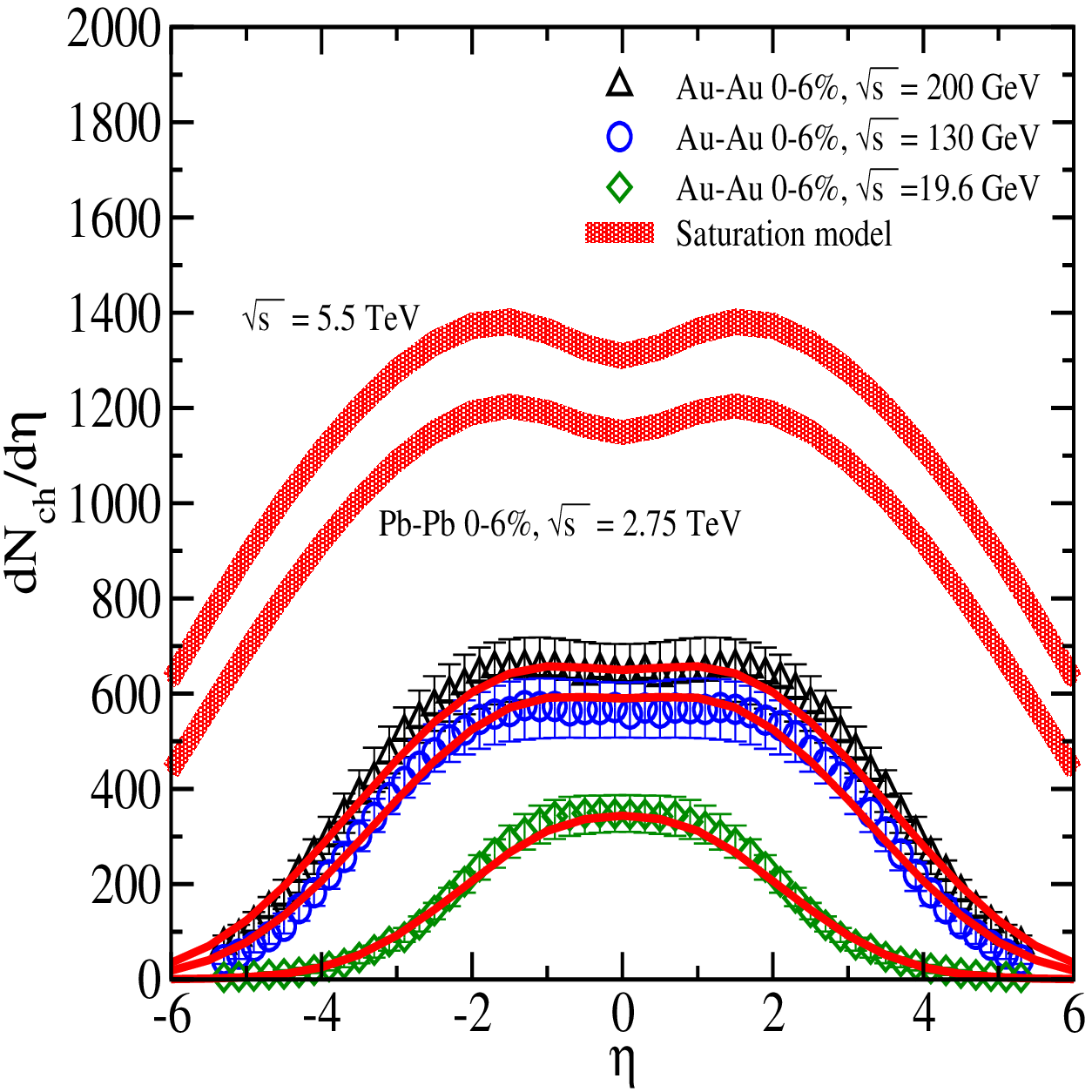} \caption{Top: The effect of the soft-scale $\mu$ is shown at $\sqrt{s}=200$ GeV.  
             Lower: Pseudo-rapidity distribution of charged particles
             produced in Au-Au and Pb-Pb central $0-6\%$ collisions at RHIC
             $\sqrt{s}=19.6, 130, 200$ GeV and the LHC energies
             $\sqrt{s}=2.75, 5.5$ TeV. The
       band indicates less than $3\%$ theoretical error coming from
       uncertainties related to normalization and modeling the mini-jet mass. The experimental data are from PHOBOS collaboration \cite{rhic1}. \label{f2}}        
\end{figure}

We have only two unknown parameters in our model:
 the overall factor $\frac{K\mathcal{C}}{\mathcal{M}}$ and the soft-scale $\mu$ introduced in the definition of mini-jet mas. We also introduced a
 $K$-factor which incorporates the discrepancy between the exact
 calculation with our formulation. These two phenomenological
 parameters are fixed at RHIC energy $\sqrt{s}=200$ GeV for Au-Au
 $0-6\%$ centrality. Then our results at other energies and various
 centralities can be considered as free-parameter predictions. The
 sensitivity of our results to various $\mu$ is shown in Fig.~\ref{f2}
 (top) at $\sqrt{s}=200$ GeV. 
 The preferred value of $\mu$ in Pb-Pb collisions is approximately of order the current-quark mass and is
 smaller than the corresponding value of $\mu\approx m_{\pi}$ in $pp$ collisions \cite{LR}. 
 The decrease in the value of the soft-scale in a denser medium is in accordance with the notion of asymptotic deconfinement that the confinement radius increases with density. 
 The preferred value of the soft-scale in $AA$ collisions is very close to what was obtained in Ref.~\cite{ris} for the value of $\Lambda_{\text{QCD}}$ at moderate densities.  
 Loosely-speaking, the smallness of $\mu$ for $AA$
 collisions compared to the $pp$ collisions is due to the fact that
 the mini-jet mass reduces in a denser medium despite the fact that the
 saturation scale increases. We checked that for Pb-Pb and $pp$
 collisions a fixed mini-jet mass about $0.14-0.2$ and $0.4-0.5$ GeV
 respectively, gives similar results to the case that the mini-jet mass runs with the saturation scale. As we already pointed out the mini-jet mass
 mimics the properties of the soft nonperturbative pre-hadronization stage and a different
 value for $\mu$ in $AA$ collisions is simply due to the fact that the
 soft pre-hadronization stage in $AA$ collisions is different from the
 corresponding one in $pp$ collisions.    

\begin{figure}[!t]
       \includegraphics[width=8 cm] {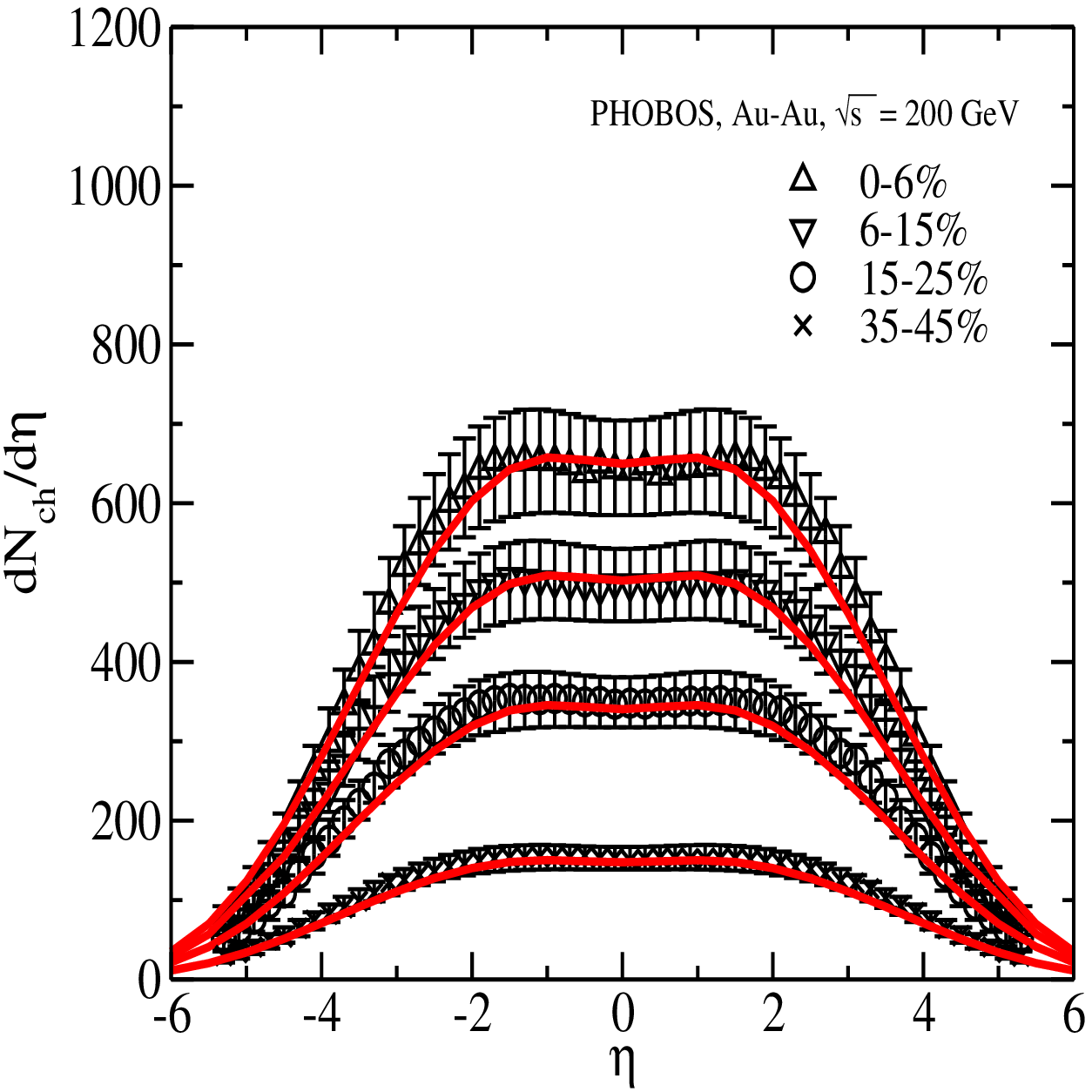}
       \includegraphics[width=8 cm] {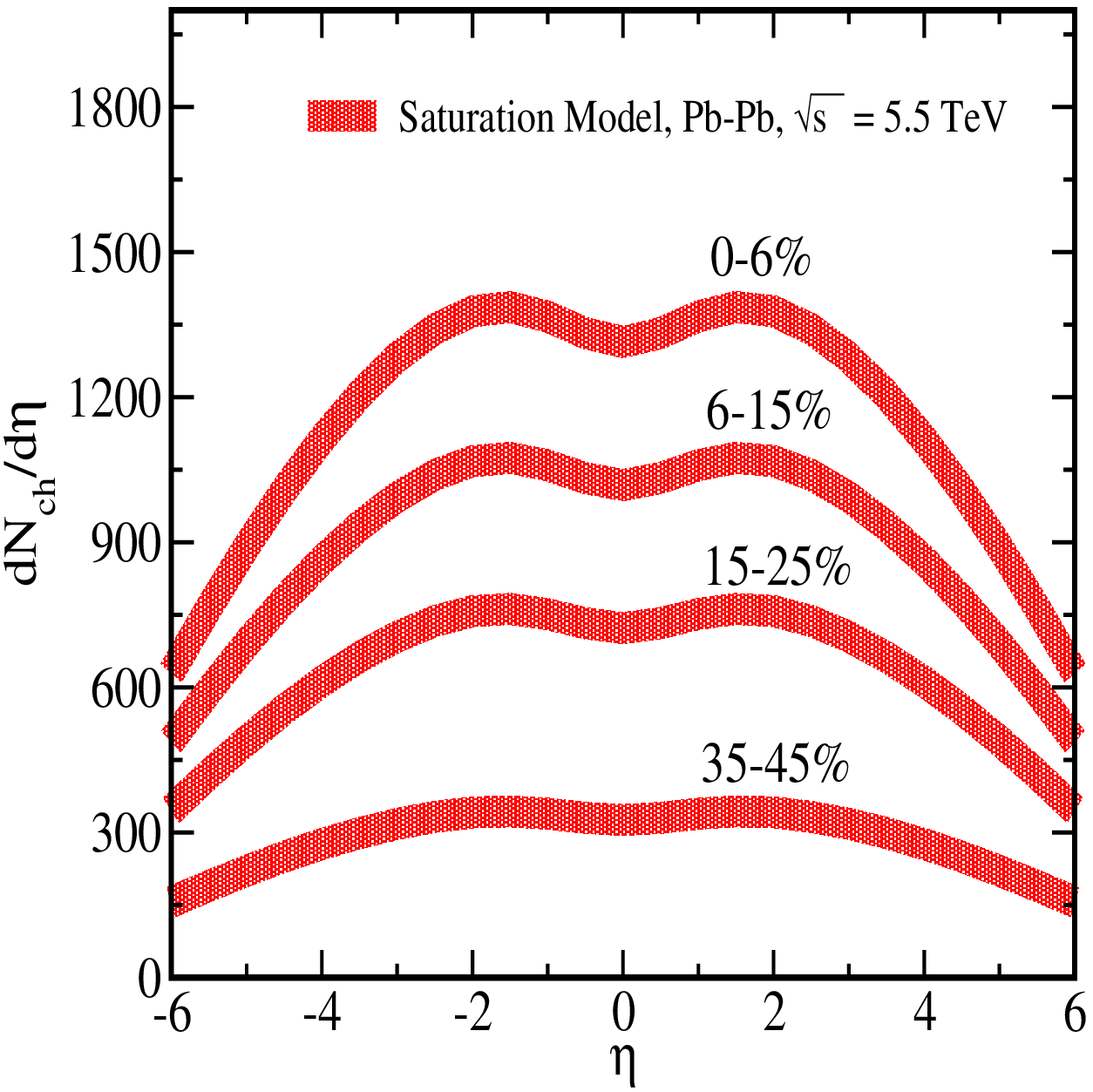} \caption{The
       pseudo-rapidity dependence at RHIC $\sqrt{s}=200$ GeV (top) and
       the LHC $\sqrt{s}=5.5$ TeV (lower) at different centrality
       bins. The band indicates less than $3\%$ theoretical errors. The
       experimental data are from PHOBOS collaboration
       \cite{rhic1}.\label{f3} }
\end{figure}

\begin{figure}[!t]
        \includegraphics[width=8 cm] {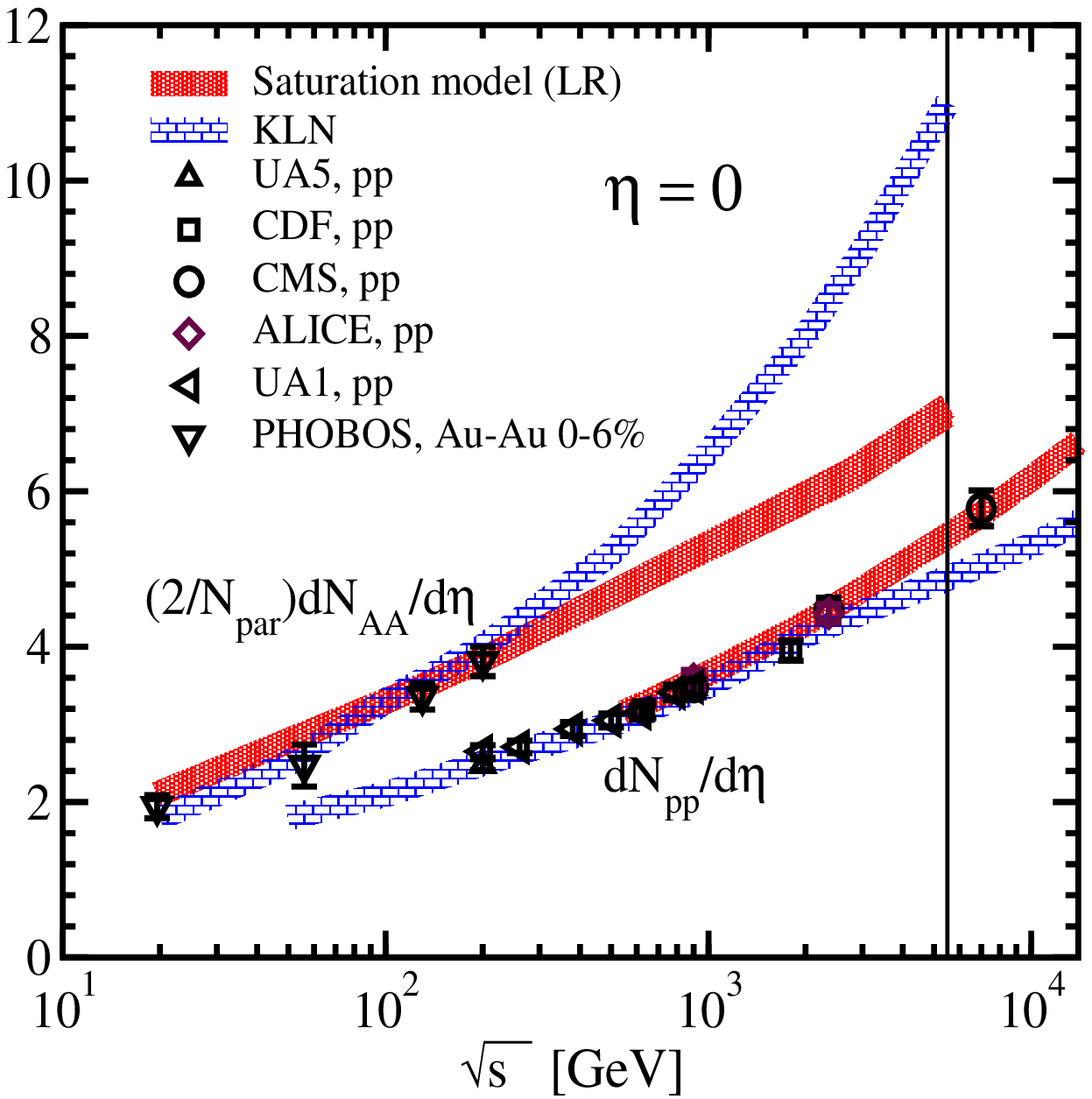} 
\includegraphics[width=8.2 cm] {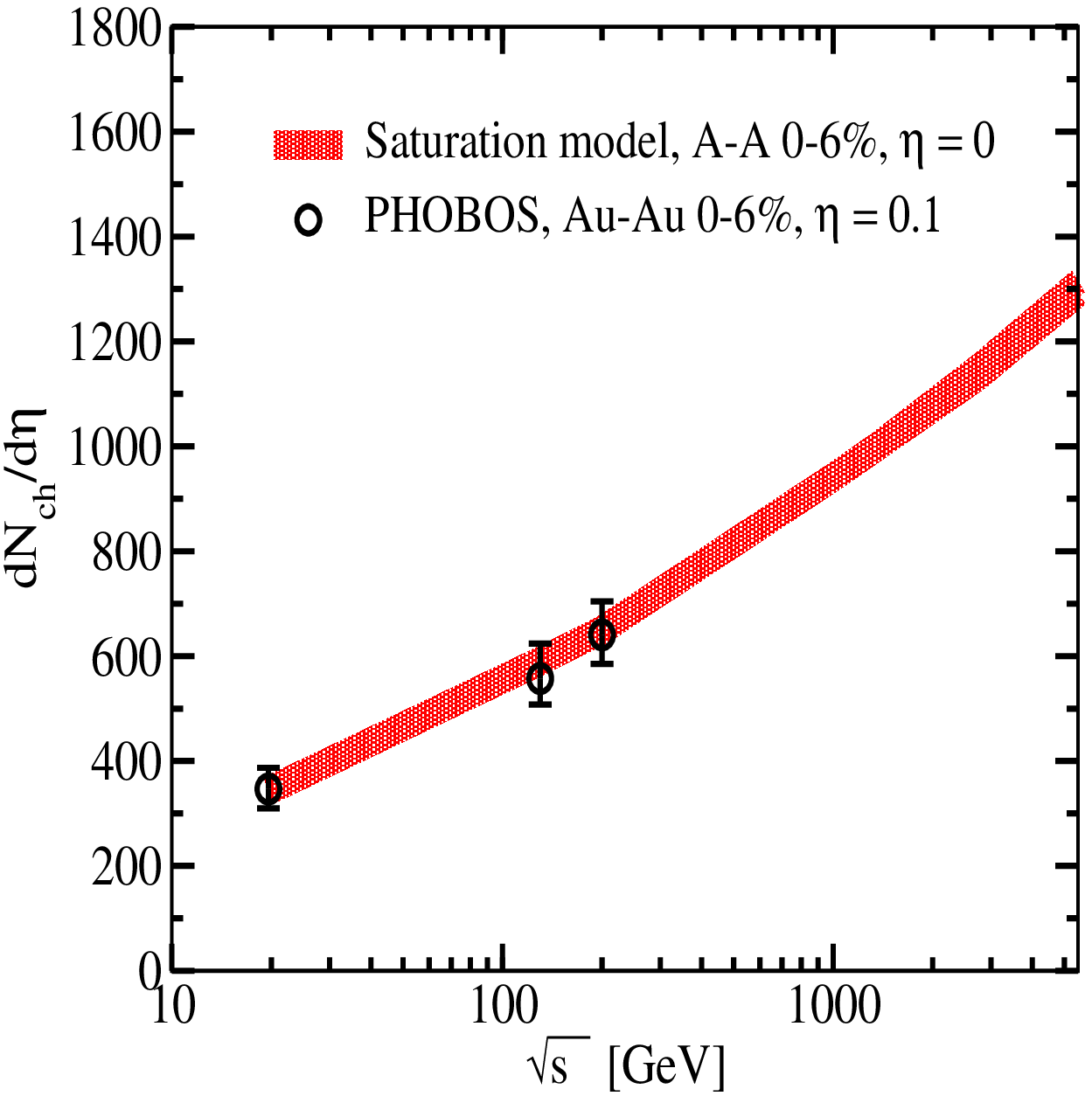} 
\caption{Energy dependence of the charged hadrons multiplicity at midrapidity $\eta=0$ in central collisions in $pp$ and $AA$ collisions.     
The theoretical curve Saturation model (LR) is our prediction. The band indicates less than $3\%$ theoretical errors. The total theoretical uncertainties is less than $7\%$, see the text for the details. We also show the KLN prediction \cite{KLNLHC} with the same error band as ours. The experimental data are from \cite{CMS,CMS2,ALICE,rhic1,ppdata}. \label{f4} }           
\end{figure}

The main sources of theoretical uncertainties in our approach stem from
assuming a fixed $\mu$ (or a fixed $m_{jet}$) and the prefactor
$\frac{K\mathcal{C}}{\mathcal{M}}$ for all energies, rapidities and
centralities. The experimental errors in the data points taken for
fixing these two unknown parameters also induce some
uncertainties. Moreover, we employed a Glauber based approach to
relate the centrality bins, number of participant and the cut in the
impact-parameters \cite{KLNLHC,WIED}. This formulation agreed with the RHIC
data on number of participant within error bars, but we expect some
uncertainties in the Glauber formulation mainly due to its
over-simplicity. In the Glauber formalism one should also assume the
value of inelastic nucleon-nucleon cross-section $\sigma_{nn}^{inel}$
(without diffractive components) from outset. Here we used the values
obtained in Ref.~\cite{GL}: $\sigma_{nn}^{inel}=64.8, 58.5, 42, 41, 30$ mb for
$\sqrt{s}=5500,2750, 200, 130, 19.6 $ GeV, respectively. Notice that
the nuclear saturation scale defined in Eq.~(\ref{M8}) can be in
principle different with exact one with extra factor. In order to
estimate this uncertainty, we checked that a factor of $2$ difference in the definition of Eq.~(\ref{M8}) (namely
$Q_A\to 2Q_A$) will change our result less than $5\%$ at the LHC. Overall, including all the above-mentioned possible theoretical uncertainties,
we expect less than $7\%$ theoretical error in our calculations at the
LHC energies. The theoretical bar in Fig.~\ref{f3},\ref{f4} only show
about $3\%$ theoretical error.

\begin{figure}[!t]
       \includegraphics[width=8 cm] {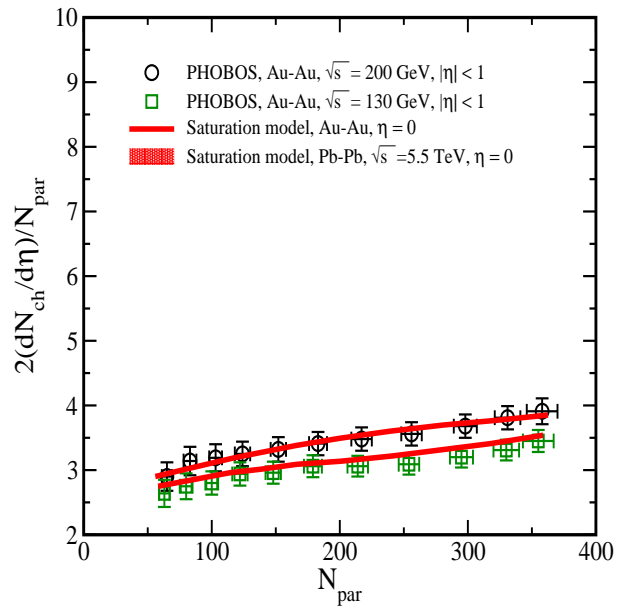}
\caption{The scaled pseudo-rapidity density as a function of number of participant $N_{par}$ at midrapidity for Au-Au at $\sqrt{s}=130, 200$ GeV and for Pb-Pb 
at $\sqrt{s}=5.5$ TeV. The experimental data are from PHOBOS collaboration \cite{rhic2}. }
\label{f5}
\end{figure}

In Fig.~\ref{f2} (lower), we show our predictions at lower RHIC
energies $\sqrt{s}=19.6$ and $130$ GeV in Au-Au collisions, and also
for the LHC energies $\sqrt{s}=2.75$ and $5.5$ TeV in Pb-Pb collisions for
$0-6\%$ centrality bin.

In Fig.~\ref{f3}, we show the charged-particle multiplicity at various
centrality bins for RHIC and the LHC high energy. As we already stressed
we only used  RHIC data at $\sqrt{s}=200$ GeV for $0-6\%$ centrality bin in order to
fix two unknown parameter of our model. Therefore, the description of other RHIC
data here can be considered as predictions. Over-all, our approach gives a very good
description of RHIC multiplicity data as a function of rapidity,
centrality, number of participant and energy, see Figs.~\ref{f2}-\ref{f5}. 

In Fig.~\ref{f4} we show the energy dependence of $dN_{pp}/d\eta$,
$dN_{AA}/d\eta$ and $(2/N_{par})dN_{AA}/d\eta$ at midrapidity $\eta=0$
for central collisions (where $N_{par}$ denotes the number of
participant for a given centrality). In Fig.~\ref{f4}, we also show
our predictions for the charged-hadron multiplicity in $pp$
collisions.  We should emphasize again that the main difference in our
formulation for the case of $pp$ and $AA$ collisions is only due to
the employed different saturation scale for a proton and a nuclear target
Eqs.~(\ref{M6},\ref{M8}) (see Fig.~\ref{f1}) and the rest of the
formulation is the same. This is in accordance with the notion of
universality of the saturation physics which can be further tested at
the LHC.  In Fig.~\ref{f4}, we show that the recent data from CMS at
$\sqrt{s}=7$ TeV remarkably confirms our predictions
\cite{LR} for $pp$ collisions. Our predictions for $dN_{AA}/d\eta$ at 
midrapidity $\eta=0$ for central Pb-Pb collisions (for $B\leq 3.7$ fm
or approximately $0-6\%$ centrality bin) for $\sqrt{s}=2.75$ and $5.5$
TeV are $1152\pm 81$ and $1314\pm 92$, respectively. 
Similar results was also suggested in Ref.~\cite{Ja} based on the numerical
solution of the BK equation with running strong-coupling with the
model assumption that the dipole amplitude does not depend on the
impact-parameter. However, in this paper \cite{Ja}, the relation between the
unintegrated gluon-density and the forward dipole-nucleus amplitude in
the $k_T$-factorization was taken a simple Fourier transformation
instead of Eqs.~(\ref{M2},\ref{M3}). The predictions of other approaches at the LHC can be found in Ref.~\cite{pr-lhc}.

In Fig.~\ref{f5}, we show $(2/N_{par})dN_{AA}/d\eta$ as a function of
number of participant for RHIC and the LHC energies at midrapidity.

 Notice that the KLN approach similar to here is also based on
 the $k_T$-factorization. Both approaches describe the RHIC data
 but provide rather different predictions at the LHC. In Fig.~\ref{f4}, we
 also show the KLN predictions for both $pp$ and $AA$
 collisions. Obviously, the KLN predictions underestimated the $pp$
 multiplicity data at the LHC $\sqrt{s}=7$ TeV (in contrast to our
 predictions) while it overestimates the multiplicity for $AA$
 collisions compared to our predictions.  The main differences between
 our approach and the KLN one is that we used explicitly the
 impact-parameter dependent form of the $k_T$-factorization and employed
 the correct relation between the unintegrated gluon density and the
 forward dipole-nucleon amplitude Eqs.~(\ref{M2},\ref{M3}). Then we
 employed an impact-parameter dependent saturation model which gives
 a good description of HERA data at small-$x$. In this sense, we did
 not have any freedom to model the saturation dipole-proton amplitude
 at RHIC or the LHC. The key difference between our approach and the KLN one is the fact 
that we rely on \eq{M8} to determine the saturation scale for nucleus while 
in the KLN approach \cite{KLNLHC} the energy dependence was used to obtain the saturation scale for nucleus at the LHC from the one at RHIC.

\section{Conclusions}
We developed a saturation approach, based on the CGC
theory, which describes DIS data at HERA \cite{WAKO}, hadron
production at the LHC \cite{LR} in $pp$ collisions and at RHIC in $AA$
collisions (this paper).  In this approach we treat in the same way
proton and nuclear target considering that the only difference between
them is the value of the saturation scale. The nonperturbative stage
of the jet-hadronization was described in the same scheme for both
proton and nuclear target, minimizing the uncertainties in
prediction that could stem from this stage. We provided here various
quantitative predictions including the rapidity, centrality and energy dependencies of the inclusive charged-hadron
multiplicity in $AA$ collisions at the LHC. We believe that our predictions for
nucleus-nucleus collisions at the LHC will be a crucial test of the CGC
approach.

\section*{Acknowledgments} 
We are thankful to Jan F. Grosse-Oetringhaus, Yuri Kovchegov and Bill Zajc for useful communication. This work
was supported in part by Conicyt Programa Bicentenario PSD-91-2006 and
the Fondecyt (Chile) grants 1090312 and 1100648.

\end{document}